\pdfoutput=1
\documentclass[journal,twoside,web]{ieeecolor}
\usepackage{tmi}
\usepackage{todonotes}
\usepackage{cite}
\usepackage{bm}
\usepackage{amsmath,amssymb,amsfonts}
\usepackage{algorithm}
\usepackage{wrapfig}
\usepackage{multirow}
\usepackage{booktabs}
\usepackage{xcolor}
\usepackage[noend]{algpseudocode}
\usepackage{amsmath}
\usepackage{graphicx}
\usepackage{textcomp}
\def\BibTeX{{\rm B\kern-.05em{\sc i\kern-.025em b}\kern-.08em
    T\kern-.1667em\lower.7ex\hbox{E}\kern-.125emX}}

\usepackage{hyperref}

\usepackage{etoolbox}
\makeatletter
\@ifundefined{color@begingroup}%
{\let\color@begingroup\relax
\let\color@endgroup\relax}{}%
\def\fix@ieeecolor@hbox#1{%
\hbox{\color@begingroup#1\color@endgroup}}
\patchcmd\@makecaption{\hbox}{\fix@ieeecolor@hbox}{}{\FAILED}
\patchcmd\@makecaption{\hbox}{\fix@ieeecolor@hbox}{}{\FAILED}
\let\originalTextcolor\textcolor
\renewcommand{\textcolor}[2]{\originalTextcolor{black}{#2}}

\begin{document}
\title{Physics-informed Score-based Diffusion Model for Limited-angle Reconstruction of Cardiac Computed Tomography}
\author{Shuo Han, Yongshun Xu, Dayang Wang, Bahareh Morovati, Li Zhou, Jonathan S. Maltz, \IEEEmembership{Senior Member, IEEE}, Ge Wang \IEEEmembership{Fellow, IEEE}, and Hengyong Yu, \IEEEmembership{Fellow, IEEE}
\thanks{The paper was submitted on %\textcolor{red}
{XXXXX}. This work was supported in part by NIH/NIBIB under grants R01EB032807 and R01EB034737, and NIH/NCI under grant R21CA264772.}
\thanks{S. Han, Y.S. Xu, D.Y. Wang, B. Morovati, L. Zhou, and H.Y. Yu are with the Department of Electrical and Computer Engineering, \textcolor{red}{University} of Massachusetts Lowell, Lowell, MA, 01854, USA.}
\thanks{J.S. Maltz is with the Molecular Imaging and Computed Tomography, GE Healthcare, Waukesha, WI, 53188, USA.}
\thanks{G. Wang is with the Department of Biomedical Engineering, \textcolor{red}{Rensselaer} Polytechnic Institute, Troy, NY, 21180, USA.}
\thanks{H.Y. Yu serves as the corresponding author (email: hengyong-yu@ieee.org).}}

\maketitle

\begin{abstract}
Cardiac computed tomography (CT) has emerged as a major imaging modality for the diagnosis and monitoring of cardiovascular diseases. High temporal resolution is essential to ensure diagnostic accuracy. Limited-angle data acquisition can reduce scan time and improve temporal resolution, but typically leads to severe image degradation and motivates for improved reconstruction techniques. In this paper, we propose a novel physics-informed score-based diffusion model (PSDM) for limited-angle reconstruction of cardiac CT. At the sampling time, we combine a data prior from a diffusion model and a model prior obtained via an iterative algorithm and Fourier fusion to further enhance the image quality. Specifically, our approach integrates the primal-dual hybrid gradient (PDHG) algorithm with score-based diffusion models, thereby enabling us to reconstruct high-quality cardiac CT images from limited-angle data. The numerical simulations and real data experiments confirm the effectiveness of our proposed approach.
\end{abstract}


\def\jsm#1{{\bf #1}}

\section{Introduction}
\label{sec:introduction}
\IEEEPARstart{C}{ardiac} Computed tomography (CT) is a powerful and widely deployed tool for non-invasive imaging of the beating heart and surrounding structures\cite{korsholm2020expert}. Specifically, electrocardiogram (ECG)-gated cardiac CT provides three-dimensional (3D) high-resolution images, enabling detailed examination of cardiac anatomy and function\cite{bharkhada2008knowledge,wang2002knowledge}. 
The evolution of CT scanners toward faster rotational times and higher tube powers is largely driven by the need to freeze the motion of the beating heart. The evolution of CT scanners toward faster rotational times and higher tube powers is largely driven by the need to freeze the motion of the beating heart. Motion artifact may affect the diagnosis of cardiovascular disease (CVD), which is the leading cause of death globally \cite{roth2020global}. The main cardiac vessels are the right coronary artery (RCA), supplying the right atrium and ventricle, the left anterior descending artery (LAD), supplying the anterior and apical left ventricle, and the left circumflex artery (LCX), supplying the lateral and posterior walls of the left ventricle. In addition, there are risks associated with the radiation dose \textcolor{red}{applied during} conventional CT imaging techniques, prompting the exploration of alternative approaches to reduce radiation exposure [3]–[5].

Limited-angle CT (LACT) imaging can potentially address these two issues by limiting the scan time, and hence the motion-sensitive time window, and patient radiation exposure. Dose reduction benefits vulnerable patient groups and patients who undergo multiple scans\cite{brenner2007dose}. 

However, the limited-angle acquisitions provide incomplete spatial frequency coverage, making it difficult to accurately reconstruct the object\cite{li2008limit}. The reconstructed images often contain streaking artifacts and distortions, particularly along directions for which data are missing. Existing strategies to enhance limited-angle CT image reconstruction quality typically utilize iterative reconstruction (IR) algorithms. These techniques optimize an objective function, which usually integrates a precise system model and an image-domain-based prior. One popularly used image prior is total variation (TV) minimization \cite{sidky2008image} and its variants \cite{niu2014sparse,zhang2021directional}. Although these IR algorithms produce images with substantially improved image quality, IR may result in the loss of fine details, and residual limited-angle-associated artifacts. In the past few years, deep learning reconstruction (DLR) has demonstrated potential to replace and supplement IR-based methods \cite{wang2020nature,morovati_reduced_2023,han2021dual,li2022physics}. DLR methodologies are widely applied in the image domain, the projection domain, and occasionally, in both. However, these methods usually require paired projection or image data which is challenging to acquire to enable supervised learning. This constraint may hinder the practicality and generalizability of DLR in certain applications.

Recently, generative models have been at the forefront of advances in computer vision. The landscape of generative models can be generally divided into two categories: likelihood-based models and implicit generative models. The former category directly approximates the data distribution. \textcolor{red}{A representative approach is the use of} variational auto-encoders (VAEs) \cite {kingma2019introduction}. The aim is to produce an output virtually identical to the input for a given input (\emph{e.g.}, an image). \textcolor{red}{VAEs can generate highly diversity} samples and easy to train. However, the fidelity of the generated samples typically remains suboptimal because the encoder predicts the distribution of the latent code and it has a pixel-based loss. In contrast, implicit generative models (\emph{e.g.}, generative adversarial networks (GANs) \cite{goodfellow2014generative}) adopt an indirect approach to fit the data distribution. These models are trained to output instances that, upon discrimination, fall within the target distribution. This strategy can generate high-fidelity samples, but the diversity of samples is \textcolor{red}{poor because adversarial loss does not incentivize covering} the entire data distribution. Notwithstanding their profound contributions to the field, implicit generative models are often criticized for their complicated training requirements, owing primarily to their dependence on adversarial learning\cite{arjovsky2016implict,wu2021bridgingimplict2}. This complexity often leads to instability and potential model collapse, posing challenges to their broad utility in real-world applications. Despite these shortcomings, both likelihood-based and implicit generative models have demonstrated great utility for many tasks in imaging.

The central problem in machine learning involves modeling complicated datasets using highly flexible probabilistic distribution families. It is challenging to ensure that learning, sampling, inference, and evaluation within these distribution families can be handled analytically or computationally. Inspired by the principles of non-equilibrium thermodynamics\cite{sohl2015deep}, diffusion models (DMs), as an alternative to the existing generative models, emerged as a potential solution for modeling the data. They demonstrate a unique and appealing approach to generative modeling\cite{ho2020denoising, song2019generative, dhariwal2021diffusion}. The basic idea of diffusion models is to systematically and gradually degrade the structures in the data distribution through an iterative forward diffusion process\cite{ho2020denoising}. This degradation is then counteracted by a reverse diffusion process that restores the structure within the data, yielding a highly flexible and easily manageable data generative model. Diffusion models have demonstrated significant advancements in imaging tasks, for example: super-resolution\cite{saharia2022image}, image in-painting\cite{kawar2022denoising}, magnetic resonance image reconstruction\cite{chung2022come}, and CT reconstruction\cite{song2021inverse}\cite{chung2023solving}. Notably, the conventional score-based diffusion models operate within the realm of unsupervised learning, offering potential advantages such as the reduction of data requirements and enhanced flexibility.

Nevertheless, in the special field of tomographic image reconstruction, there is an underlying desire for the ability to manipulate the final outcome to not only include image denoising and artifact removal but also to ensure the maintenance of image quality, while enhancing \textcolor{red}{imaging metrics such as} contrast resolution. Our results suggest that the proposed method can achieve high-quality reconstructions from limited-angle data. It appears to address the issue of false structure generation, which could enhance the reliability and clinical applicability of the reconstructed images. Primal-dual hybrid gradient (PDHG) is often more robust to noise and other perturbations than ADMM\cite{komodakisadmm}. This can be beneficial in limited-angle CT reconstruction, where the data is often noisy and under-sampled.

In the pursuit of high-quality image reconstruction for limited-angle CT imaging, we introduce a novel approach we term a physics-informed score-based diffusion model (PSDM). This approach combines a model-based optimization strategy with the sampling steps of the score-based model,  allowing simultaneous utilization of both data-based (as in DLR) and model-based (as in IR) priors. First, the score-based models are employed to generate an initial image, leveraging a score. Subsequently, we combine the initial image and the limited-angle CT image in the Fourier domain to leverage the complementary frequency domain characteristics. Finally, another data consistency step is implemented via the PDHG optimization. This step is instrumental in eliminating the discrepancy between the initial and target images. Furthermore, isotropic TV regularization is utilized to enhance the refinement of the final image.

\section{Background}
\subsection{CT imaging model}
CT acquisition is primarily affected by two kinds of noise. The first type is \textcolor{red}{electronic} noise, which originates from the detection system and adheres to a Gaussian distribution, being independent of object structure. The second type is photon statistical noise from inherent quantum fluctuations during X-ray emission, and it usually conforms to a \textcolor{red}{Poisson distribution, the mean of which, for a particular ray, is dependent on the x-ray source beam characteristics and the structure of the imaged object.} In the projection domain, the noise exhibits a complicated statistical distribution that blends both Gaussian and Poisson noise:
\begin{equation}
    \boldsymbol{n} \thicksim  \mathcal{N}(0,\sigma^2) +  \mathcal{P} (\xi(\boldsymbol{x})) ,
\end{equation}
where $\boldsymbol{n}$ is the pre-log noise variable, $\mathcal{N}(0,\sigma^2)$ denotes the Gaussian distribution \textcolor{red}{with variance} $\sigma^2$, and $\mathcal{P} (\xi(\boldsymbol{x}))$ is Poisson distribution in which  $\xi(\boldsymbol{x})$ denotes the latent mapping from image structure $\boldsymbol{x}$ to the distribution \textcolor{red}{expectation}.
\textcolor{red}{In a general model of CT image reconstruction $\boldsymbol{x}$ must fulfill:}
\begin{equation}
\boldsymbol{y} = \boldsymbol{A} \boldsymbol{x} + \boldsymbol{\tilde{n}} ,
\label{eq1}\end{equation}
where $\boldsymbol{y} \in \mathbb{R}^M$ is a vectorized sinogram, $\boldsymbol{x} \in \mathbb{R}^N$ is a vectorized CT image, $\boldsymbol{\tilde{n}} \in \mathbb{R}^M$  is the post-log noise corresponding to the pre-log noise $\boldsymbol{n}$,  $\boldsymbol{A} \in \mathbb{R}^{M \times N}$ is a CT system matrix, and $M$ and $N$ are respectively the total numbers of measurements in the sinogram and pixels in the image domain. Because the limited-angle CT reconstruction problem is ill-posed, a standard approach to estimate the unknown image $\boldsymbol{x}$ is to employ a regularization term:
\begin{equation}
\min_{x} \frac{1}{2} \| \boldsymbol{A}\boldsymbol{x} - \boldsymbol{y} \|_{2}^{2} + \lambda R(\boldsymbol{x}).
\label{eq2}\end{equation}
The first term of Eq.(\ref{eq2}) is a data fidelity term based on the $L_2$-norm. The second term, $R(\boldsymbol{x})$, is the regularizer that incorporates the prior information of the solution, and $\lambda$ is a trade-off parameter to balance these two terms.

\def\xT{\boldsymbol{x}_{T}}
\def\xt{\boldsymbol{x}_{t}}

\begin{figure*}[!t]
\centering
\includegraphics[width=0.7\textwidth]{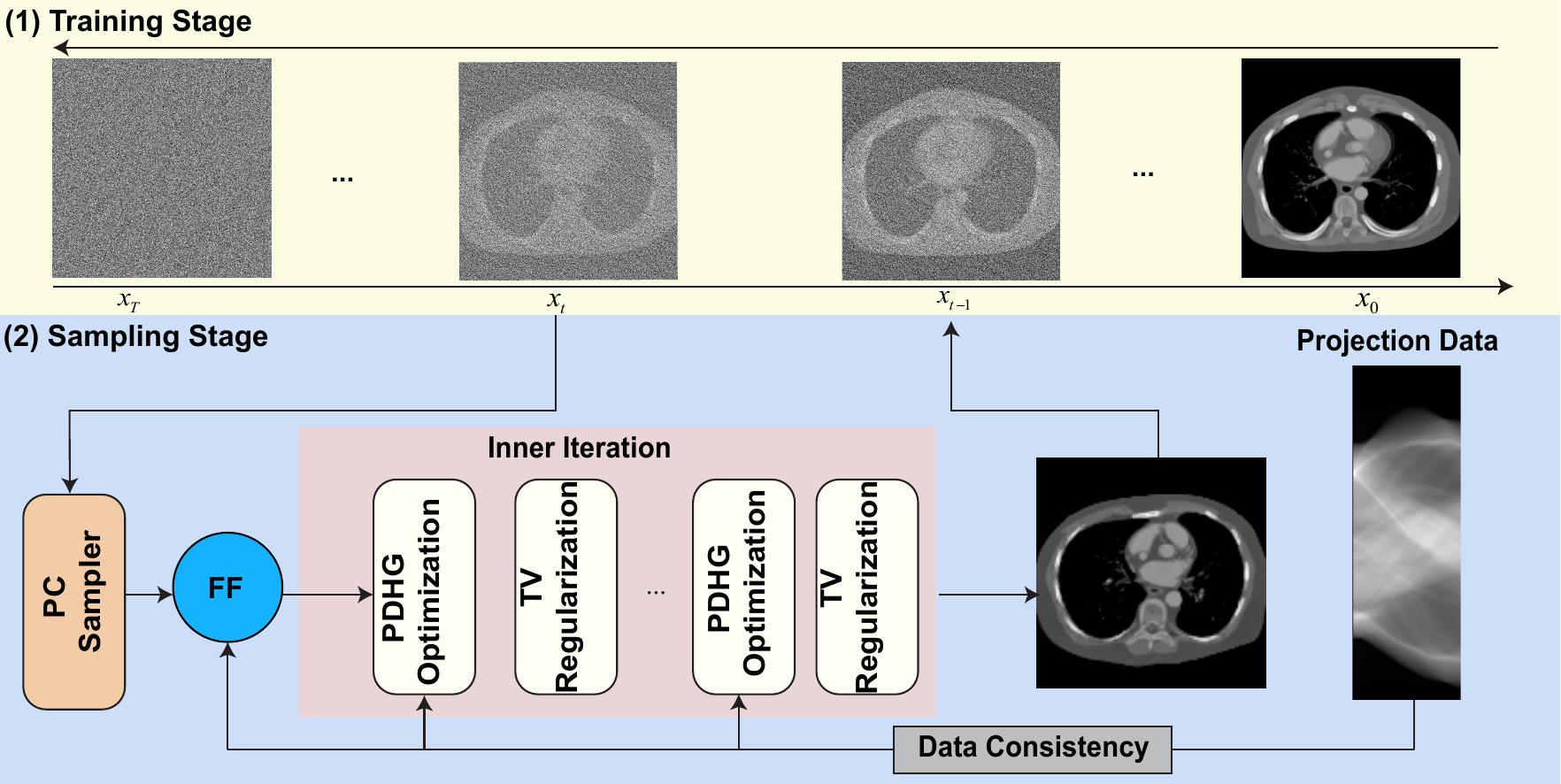}
\caption{Architecture of the PSDM image reconstruction framework. In the initial phase, a score-based diffusion model is employed to generate a preliminary image. This is followed by the implementation of a data consistency constraint, which integrates Fourier fusion and the \textcolor{red}{primal-dual hybrid gradient} (PDHG) optimization technique. Subsequently, a data consistency constraint including Fourier Fusion(FF) and PDHG optimization is introduced. By alternatively performing these two processes, high-quality reconstruction result \textcolor{red}{may be achieved.}}
\label{Workflow}
\end{figure*}

\subsection{Score-based Diffusion Model \& SDEs}
Score-based diffusion models are generative models in which samples from a probability density $p_{\text{data}}(\boldsymbol{x})$ can be produced via Langevin dynamics, a stochastic process used for sampling from probability distributions. In a score-based diffusion model, Langevin dynamics can be conceptualized as a methodical process for generating data points that are statistically consistent with the original dataset. It commences with an arbitrary data point and employs iterative updates to gradually align this point with the characteristics of the dataset. Central to this mechanism is the score function, $\nabla_{\boldsymbol{x}} \log p(\boldsymbol{x})$\cite{song2019generative}, which represents the gradient of the logarithm of the probability density. This function provides directional guidance, indicating the path towards regions with higher data point probabilities. By adhering to this guidance, Langevin dynamics effectively moves the initial arbitrary point towards areas more densely populated by actual data points.

More specifically, given a fixed step size $\delta > 0$, and an initial value $\boldsymbol{x}_{0}$ (sampled from a prior distribution), the Langevin dynamics sampling process \cite{song2019generative} can be modelled as:
\begin{equation}
\boldsymbol{x}_{t} = \boldsymbol{x}_{t-1} + \frac{\delta}{2} \nabla_{\boldsymbol{x}} \log p(\boldsymbol{x}_{t-1}) + \sqrt{\delta} N_{t},
\label{Langevin}\end{equation}
where $t$ represent the sampling step, and  $N_{t} \sim \mathcal{N}(0,\boldsymbol{I})$ denotes random disturbance, where $\boldsymbol{I}$ \textcolor{red}{is the} identity matrix. The distribution of $\xT$ equals $p_{\text{data}}(\boldsymbol{x})$ when $\delta \rightarrow 0$ and the number of sampling steps $T \rightarrow \infty$.
From Eq.(\ref{Langevin}), to obtain samples from $p_{\text{data}}(\boldsymbol{x})$, we need to train a score network $\boldsymbol{s}_{\boldsymbol{\theta}}(\boldsymbol{x})$ to estimate $\nabla_{\boldsymbol{x}} \log p_{\text{data}}(\boldsymbol{x})$. This is the basic idea of score-based diffusion model.

In both vanilla\cite{ho2020denoising} and score-based\cite{song2019generative} diffusion models, the diffusion and sampling processes are manually divided into T fixed steps. However, in reality, the diffusion and sampling procedures should not be deliberately segmented into steps. Instead, each can be \textcolor{red}{considered} as a continuous transformation process over time, characterized by stochastic differential equations (SDEs)\cite{song2020sde}. This provides a generalized representation by utilizing continuous processes to describe diffusion models. The diffusion process using SDEs can be expressed as\cite{song2020sde}:
\begin{equation}
d \boldsymbol{x}=\boldsymbol{f}(\boldsymbol{x}, t) \, dt+g(t) \,d \boldsymbol{w},
\label{sde_forward}\end{equation}
where $\boldsymbol{f}(\boldsymbol{x}, t)$ is called drift coefficient \textcolor{red}{and represents} the deterministic process,  $g(t)$ is the diffusion coefficient \textcolor{red}{that introduces} the stochastic process, and $d\boldsymbol{w}$ represents the derivative of Wiener Process \textcolor{red}{and introduces} perturbations at any point in time and \textcolor{red}{induces} uncertainty. $\boldsymbol{f}(\boldsymbol{x}, t)$ is a function of the data samples $\boldsymbol{x}$ and time $t$, and is deterministic. Both  terms in Eq.(\ref{sde_forward}) are essential to score-based diffusion models: one drives the system towards the high-density distribution of the data, and the other introduces randomness to ensure diversity among the generated samples. We note that $t$ in Eq.(\ref{sde_forward}) represent a continuous time variable, whereas it \textcolor{red}{is} a discrete time variable in Eq.(\ref{Langevin}). We disambiguate this notation only when necessary.

Among the possible choices of $\boldsymbol{f}$ and $g$, we choose the variance-exploding SDE (VE-SDE)\cite{song2020sde}. Let $\boldsymbol{f}(\boldsymbol{x}, t) = 0$ and $g(t) = \sqrt{\frac{d[\sigma^2(t)]}{dt}}$. We can simplify the forward process into a Wiener process:
%\jsm{don't know how to fix this. We specifiy the forward process as a Wiener process? ans:yes}
\begin{equation}
d\bm{x} = \sqrt{\frac{d[\sigma^2(t)]}{dt}}\,d\bm{w},
\label{vesde}\end{equation}
where $\sigma^2(t)$ refers to the time-dependent variance function. Because $\sigma^2(t)$ is designed to follow a rapidly increasing variance schedule, its derivative is guaranteed to be non-negative.
In addition to the forward diffusion process, we also need consider the reverse sampling process in which time $t$ regresses from 1 to 0. In this reverse process, the goal is to transform a simple Gaussian distribution $\mathcal{N}(0,\boldsymbol{I})$ back into the desired data distribution $p_{\text{data}}(\boldsymbol{x})$. By applying the Anderson's theorem\cite{anderson1982reverse} to (\ref{vesde}), we have the reverse SDE:
\begin{equation}
    d \boldsymbol{x}=-\frac{d\left[\sigma^2(t)\right]}{d t} \nabla_{\boldsymbol{x}} \log {p}_t\left(\boldsymbol{x}\right) d t+\sqrt{\frac{d\left[\sigma^2(t)\right]}{d t}} d \overline{\boldsymbol{w}}.
\label{reverSDE}\end{equation}
In Eq.(\ref{reverSDE}), $\overline{\boldsymbol{w}}$ denotes the reverse Wiener process. It is essential to observe that the drift term is defined by the score function $\nabla_{\boldsymbol{x}} \log {p}_t\left(\boldsymbol{x}\right)$. Hence, the problem of reverse sampling from the data distribution is transformed into a problem of score estimation. However, direct estimation of the score function of high-dimensional distributions is challenging. The denoising score matching (DSM) method allows us to estimate the score function using a denoising function. Following the DSM framework, we introduce a denoising function: 
\begin{equation}
\min_{\theta} \mathbb{E}_{t,\boldsymbol{x}(t)} \bigl[ \lambda(t) \, \left\| \boldsymbol{s}_{\boldsymbol{\theta}}(\boldsymbol{x}(t), t) - \nabla_{\boldsymbol{x}_t} \log p(\boldsymbol{x}(t)|\boldsymbol{x}(0)) \right\|_2^2 \bigr].
\label{DSM_loss}
\end{equation}

In this loss function, $\boldsymbol{s}_{\boldsymbol{\theta}}(\boldsymbol{x}(t), t)$ is the score function estimated by a model (\emph{i.e.}, U-Net) parameterized by $\boldsymbol{\theta}$, $\nabla{\boldsymbol{x}_t}\log p(\boldsymbol{x}(t)|\boldsymbol{x}(0))$ is the true score function of the data distribution, and $\lambda(t)$ is a weighting factor that may vary depending on the time $t$.
The denoising function has the desirable property that its expectation equals the score of the perturbed data at the input location. This makes it possible to replace the score function in the reverse SDE with the denoising function. This property implies that we can approximate the reverse SDE with the following equation:
\begin{equation}
    d \boldsymbol{x}=-\frac{d\left[\sigma^2(t)\right]}{d t} \,\boldsymbol{s}_{\boldsymbol{\theta}}\left(\boldsymbol{x}, t\right) d t+\sqrt{\frac{d\left[\sigma^2(t)\right]}{d t}} \, d \overline{\boldsymbol{w}}.
\label{reverseSDE_approx}\end{equation}

By doing so, we simplify the reverse process by directly estimating the denoising function, instead of the complex score function. A neural network can be trained to model $\boldsymbol{s}_{\boldsymbol{\theta}}\left(\boldsymbol{x}, t\right)$, with the training objective being a regularized version of the DSM loss function.
In practice, we can leverage the variance of the VE-SDE to control the amount of noise injection. Hence, the reverse process is trained to denoise observations at different noise levels, resulting in a more robust score estimation. The model's ability to handle different noise levels leads to an improvement in the quality of samples drawn from the learned data distribution.

In conclusion, the approach based on VE-SDE and DSM provides a robust and effective way to estimate the score function of complex and high-dimensional distributions, enabling the transformation of a simple Gaussian distribution into desired data distribution. The predictor-corrector (PC) sampler\cite{song2020sde} can be used to solve Eq.(\ref{reverseSDE_approx}) by discretization of the time interval [0, 1] into $\mathit{I}$ steps. In our case $\mathit{I}$ is set to 1000.

\subsection{Conditional predictor-corrector sampler}
Unconditional generation often explores upper limits in terms of effects, while conditional generation mainly focuses on application-oriented tasks because it allows us to control output results according to specific requirements (\emph{i.e.,} generation of high-quality CT images using limited-angle projections). To address inverse problems in CT image reconstruction, conditional generation is indispensable. Generally speaking, there are two types of strategies for conditional control generation: classifier-guidance\cite{dhariwal2021diffusion} and classifier-free\cite{ho2021classifier}. For CT reconstruction, the classifier-guidance approach allows the model to leverage an angular classifier that provides probabilistic conditioning guidance based on the view angle of the limited projections. This enables tighter conditioning on the measurement data, and is less computationally expensive than training a diffusion model. Hence, it is more convenient for subsequent researchers to fine-tune a pretrained model based on rich classifier gradients, thereby \textcolor{red}{obviating the need to train a new model.}

For the case of limited-angle CT reconstruction, we wish to sample from a posterior distribution. This can be defined as the conditional distribution of $\boldsymbol{x}$ given the measurement $\boldsymbol{y}$, \emph{i.e.,} $p(\boldsymbol{x}|\boldsymbol{y})$. According to Bayes' theorem, the score function of the posterior distribution can be computed as:
\begin{equation}
\nabla_{\boldsymbol{x}}\log p_t(\boldsymbol{x}|\boldsymbol{y}) = \nabla_{\boldsymbol{x}} \log p_t(\boldsymbol{x}) + \nabla_{\boldsymbol{x}} \log p_t(\boldsymbol{y}|\boldsymbol{x}).
\label{likeli}\end{equation}
Here, the likelihood enforces the data consistency, and thereby \textcolor{red}{induces} samples that satisfy $\boldsymbol{y}=\boldsymbol{A} \boldsymbol{x}$. The calculation of Eq.(\ref{likeli}) is decomposed into two stages: the initial stage consists of a denoising step, where the previously acquired data prior is applied to denoise the image, so we can simply use the pre-trained score function $\boldsymbol{s}_{\boldsymbol{\theta}}$. The second stage is projection onto the measurement subspace.
Within a discrete framework, this process is:
\begin{equation}
\boldsymbol{x}_{i-1}^{\prime} \leftarrow \text{solve}(\boldsymbol x_{i-1}, \boldsymbol{s}_{\boldsymbol{\theta}}),
\label{solve1}\end{equation}
\begin{equation}
\boldsymbol x_{i} \leftarrow P \{x|Ax=y\} (\boldsymbol{x}_{i-1}^{\prime}),
\label{solve2}\end{equation}
where "solve" represents a numerical solver of Eq. 9, and $P$ denotes the projection operator by following the \textcolor{red}{Projection Onto Convex Sets(POCS)} principle to satisfy the data constraint. In our case, we use the PC sampler as a solver for Eq.(\ref{solve1}) and the PDHG as a data-consistency module to solve Eq.(\ref{solve2}). It should be noted that we directly applied our PDHG optimization steps to the noisy variables in Eq.\ref{solve2} by following \cite{chung2023solving,chung2022come, chung2022score}. This works when we use sufficiently small step sizes to enable correction along the way. The overall framework of our proposed method \textcolor{red}{appears} in Fig. 1. We use a trained network $\boldsymbol{s}_{\boldsymbol{\theta}}$ to denoise and use PDHG to enforce the data consistency. 

The PC sampler, inspired by the predictor-corrector method in numerical solutions for differential equations, is utilized in the reverse process to generate new samples.
Unlike traditional SDE samplers such as Euler-Maruyama (EM) and Runge-Kutta (RK) \cite{Kloeden2013}, this sampling strategy enhances the final solutions by incorporating additional information. The process involves predicting the next state in the diffusion process (the ``predictor" step), and then correcting this prediction based on the actual observed data (the ``corrector" step). In summary, the PC sampler alternates between two steps:
\begin{enumerate}
    \item Predictor step: Given the current state of the sample, it predicts the next state by taking a step in the direction of the gradient of the log probability of the data. This is equivalent to performing a step of gradient ascent on the log probability.
    \item Corrector step: It corrects the prediction by adding a small amount of noise. This noise is sampled from a Gaussian distribution whose standard deviation is determined by the diffusion process.
\end{enumerate}
By alternating between the above two steps, the PC sampler gradually transforms a sample in the prior distribution into a sample in the data distribution, enabling the score-based diffusion model to generate new samples that closely resemble the training data.

\section{Methodology}
\subsection{Motivation}
While the utilization of a score-based diffusion model is beneficial, it is important to acknowledge its potential limitations. Specifically, there is a risk of generating false structures or artificial elements that do not correspond to the true underlying structures in the image. This is a consequence of employing a model that is entirely data-driven. Such a model relies heavily on the quality and representativeness of the input data. 

To overcome this issue, we incorporate physical constraints. Physical constraints are derived from established theories or empirical observations that guide us on how the data should behave, providing an additional layer of control and validation on the output of the diffusion model. This can help to correct for aberrations such as false structures, aligning the output more closely to the known physical truths and thus ensuring a more accurate and robust image representation. The combination of data prior and model has proven an accurate and stable method \textcolor{red}{in image reconstruction applications}\cite{wu2022stabilizing1}. To combine model priors from the physics models and data priors from the diffusion model, the projection operator $P$ in Eq.(\ref{solve2}) is replaced by the PDHG update method, and the Fourier fusion strategy is used between diffusion model denoised result and limited-angle CT image.
\subsubsection{PDHG}
The rationale for employing the PDHG methodology is well-known in medical image processing, and the minimization process in Eq.(\ref{eq2}) can be viewed as an optimization task. Because the objective function \textcolor{red}{is} usually is non-differentiable, we are unable to utilize the gradient descent approach for optimization. An alternative strategy is to apply non-smooth convex optimization techniques. Proximal methods can work directly with non-smooth objective functions. In this case, a proximal step takes the place of the gradient step. In a proximal primal-dual scheme, an additional dual variable that is in the range of the operator is introduced, and the primal and dual variables are updated alternately. For the optimization problem below, three vector spaces are used: $I$ is the space of 2-dimensional (2D) discrete images, $D$ is the space of the CT projection data, and $V$ is the space of spatial-vector-valued image arrays ($V = I^2$). We define the indicator function:
\begin{equation}
\label{Boxindicator}
\delta_{\textrm{Box}(\alpha)}(x) \equiv
\begin{cases}
0 & \|x\|_\infty \le \alpha \\
\infty & \|x\|_\infty > \alpha
\end{cases},
\end{equation}
where \(\| \cdot \|_\infty\) is an \(\infty\)-norm used to select the maximum absolute value of the components of a vector. The set \(\text{Box}(\alpha)\) contains vectors in which no component is greater than \(\alpha\). This means that all components of a vector in this set lie in the interval \([- \alpha, \alpha]\). In 2D, \(\text{Box}(\alpha)\) is a square centered at the origin with side length \(2\alpha\); \textcolor{red}{for any} point \((x,y)\) inside this square, both \(x\) and \(y\) must lie within the interval \([- \alpha, \alpha]\). Additionally, in the context of a given vector space \(X\), the notations \(\mathbf{0}_X\) and \(\mathbf{1}_X\) are employed to denote vectors in \(X\) whose components are all set to 0 and 1, respectively. For instance, in a 2D space, \(\mathbf{0}_X\) \textcolor{red}{is} \([0,0]\) and \(\mathbf{1}_X\) \textcolor{red}{is} \([1,1]\).
\subsubsection{Fourier Fusion}
Fourier fusion offers a promising solution to solving limited-angle artifacts by leveraging the complementary strengths of diffusion models and Fourier domain processing. \textcolor{red}{Diffusion models are} good at denoising and generating high-quality images from noisy or incomplete data. By integrating Fourier fusion into the framework, we aim to enhance spatial frequency components that are underrepresented or distorted due to the generative process from diffusion model, thus improving the overall resolution and detail in the reconstructed image. This approach utilizes a region combination strategy in the Fourier domain, where specific frequency regions from each source are selectively merged to optimize image quality.

To ensure coherence and to maximize the use of high-quality information from the limited-angle CT image ($\boldsymbol{X}_{\text{LACT}}$ which obtained using FBP algorithm), a unified mask $M$ is adopted. $M$ is constructed based on the centered Fourier transform $\mathcal{F}$ (first applying the Fourier transform and then centering the spectrum) of the limited-angle CT image. This mask identifies frequency components that are less affected by the limitations of \textcolor{red}{the set of} acquisition angle and set those areas to zero. Those frequency components represent reliable information about the object's structure. The final step in the reconstruction process involves the inverse centered Fourier transform $\mathcal{F}^{-1}$ (first applying the inverse shift and then \textcolor{red}{applying} the inverse transform) of combined image to convert it from the frequency domain to the spatial domain, thereby producing the fused image.

\subsection{Algorithmic Steps}
A well-known primal-dual scheme is the PDHG algorithm, also referred to as the Chambolle-Pock algorithm\cite{chambolle2011first}. It applies to a general form of the primal minimization and dual maximization:
\begin{equation}
\underset{x \in X}{\text{min}} \left( F(K(x)) + G(x) \right) ,
\end{equation}
\begin{equation}
\underset{y \in Y}{\text{max}} \left(-F^{*}(y) - G^{*}(-K^{T}y) \right).
\end{equation}
In this context, \(x\) and \(y\) denote vectors of finite dimensionality within the spaces \(X\) and \(Y\), respectively. The symbol \(K\) is a linear operator mapping from \(X\) to \(Y\). The functions \(G\) and \(F\), which may not be smooth, represent convex mappings from the respective spaces \(X\) and \(Y\) to a set of non-negative real numbers. The superscript ``\(\ast\)" represents convex conjugation.

Particularly, in the context of Eq.(\ref{eq2}), we apply an isotropic TV semi-norm as a regularization term denoted as $R(x) = \|(\nabla |x|)\|_1$. This norm has \textcolor{red}{proven} useful for edge-preserving regularization in the PDHG optimization algorithm\cite{sidky2012convex}. The spatial-vector depiction $\nabla x$ corresponds to a discrete approximation of the image gradient, residing in the vector space $V$. As reported in \cite{sidky2012convex}, the objective function of PDHG to solve Eq.(\ref{eq2}) can be derived as the following formulas:
\begin{equation}
\begin{aligned}
& F(x, z)=F_1(x)+F_2(z), \\
& F_1(x)=\frac{1}{2}\|m-y\|_2^2, \quad F_2(z)=\lambda\|(|z|)\|_1, \\
& G(x)=0, \\
& m=Ax,  z = \nabla x, \\
& K=\left(\begin{array}{l}
A \\
\nabla
\end{array}\right),
\end{aligned}
\end{equation}
where $x \in I$, $y \in D$ and $z \in V$. To obtain the convex conjugate of $F$, variables $p$, $q$ and $r$ are introduced as auxiliary components to play a crucial role in the computation of the Lagrange multipliers. Then, we have the convex conjugate of $F$ and $G$. Calculating the convex conjugate is essential for transitioning between the primal and dual spaces, thereby facilitating the primal-dual updates in the PDHG algorithm:
\begin{equation}
F^*(p,q) =\frac{1}{2}\|p\|_2^2+\langle p, y\rangle_D + \delta_{\mathrm{Box}(\lambda)}(|q|), 
\end{equation}
\begin{equation}
G^*(r) =\delta_{\mathbf{0}_I}(r),
\end{equation}
where $p \in D$, $q \in V$ and $r \in I$.

Finally, we obtain the proximal mapping of the PDHG algorithm for Eq.(\ref{eq2}):
\begin{equation}
\operatorname{prox}_\sigma\left[F^*\right](y, z)=\left(\frac{m-\sigma y}{1+\sigma}, \frac{\lambda z}{\max \left(\lambda \mathbf{1}_I,|z|\right)}\right),
\end{equation}
where $\sigma$ is the step size to control the update of the variables in each iteration, and $\mathbf{1}_I$ is an image with all pixels set to 1. The proximal mapping function in this context serves as a mechanism to iteratively update the dual variables \(y\) and \(z\), thereby influencing the optimization of the primal variable \(x\) in the PDHG algorithm, ensuring convergence to an optimal solution that balances data fidelity and regularization.

Consequently, we can establish our PSDM algorithm as \textbf{Algorithm 1}. It should be noted that there are two iterative steps [see Eqs.(\ref{step1}) and (\ref{step2})]: 1) The denoising step of SDE indexed by $i$ , and 2) PDHG iteration indexed by $n$. In our case $n$ is set to 30. We apply the Fourier fusion and then in the second iteration step and initialization the $\boldsymbol{x}$ = $\boldsymbol{x}'_i$.
\begin{equation}
\boldsymbol{x}_{\text{DN}} \gets\text{solve}(\boldsymbol{x}_{i}, \boldsymbol{s}_{\boldsymbol{\theta}}),
\label{step1}\end{equation}

\begin{equation}
    \boldsymbol{x}'_i \gets \mathcal{F}^{-1}\left\{M \left[ \mathcal{F}(\boldsymbol{x}_{\text{DN}})\right] + \mathcal{F}(\boldsymbol{X}_{\text{LACT}}) \right\},
\label{FF}\end{equation}

\begin{equation}
\boldsymbol{x}_{i-1} \leftarrow \underset{\boldsymbol{x}}{\operatorname{argmin}} \frac{1}{2}\left\|\boldsymbol{y}-\boldsymbol{A} \boldsymbol{x}\right\|_2^2+\|(|\nabla \boldsymbol{x}|)\|_1.
\label{step2}\end{equation}
The purpose of initial iteration is to transform the dataset into the learned data distribution. The succeeding iteration \textcolor{red}{can} be interpreted as a data fidelity term to ensure the congruity between the generated data and the provided input. Specifically, Fourier fusion is strategically applied between steps 400 to 800 of the sampling process. This timing is chosen because the initial image quality at the very beginning is too poor for Fourier fusion to be effective. Additionally, introducing Fourier fusion too late in the process, particularly at the final stages, \textcolor{red}{can potentially negatively} impact the quality of the final reconstructed image.
\begin{algorithm}
\caption{Pseudocode of the PSDM for sampling. The constant $L$ is the norm of the combined linear transform$\|(A,\nabla)\|_2$, and $\tau$ and $\sigma$ are non-negative PDHG algorithm parameters which are both set to be $\frac{1}{L}$. $\theta \in \left[ 0 , 1 \right] $ is another PDHG algorithm parameter which is set to be 1. And $\boldsymbol{s}_{\boldsymbol{\theta}}$ represents trained neural network model parameters.}
\begin{algorithmic}[1]
\Require $\nabla_{\boldsymbol{x}} \log p_t(\boldsymbol{x})$, $N$, $I$, $\lambda$
\State Initialization $x_N \sim \mathcal{N}(0, \sigma_T^2 I)$, $L \gets \|(\boldsymbol{A},\nabla)\|_2$, $\tau \gets \frac{1}{L}$, $\sigma \gets \frac{1}{L}$, $\theta \gets 1$;
\For{$i=I-1$ \textbf{to} $0$}
\State $\boldsymbol{x}_{\text{DN}} \gets \text{solve}(x_i, \boldsymbol{s}_{\boldsymbol{\theta}})$ \Comment{SDE Denoising}
\State $\boldsymbol{x}'_i \gets \mathcal{F}^{-1}\left\{M \left[ \mathcal{F}(\boldsymbol{x}_{\text{DN}})\right]+ \mathcal{F}(\boldsymbol{X}_{\text{LACT}}) \right\}$ \Comment{FF}
\State Initialization $u_0, p_0, q_0, n \gets 0, x_n \gets 0$, $\boldsymbol{x}'_n \gets \boldsymbol{x}'_i$
\Repeat
\State $p_{n+1} \gets (p_n + \sigma(\boldsymbol{A}\boldsymbol{x}'_n - y))/{(1 + \sigma)}$
\State $q_{n+1} \gets \lambda (q_n + \sigma\nabla \boldsymbol{x}'_n)/{\max(\lambda \mathbf{1}_I,|q_n + \sigma\nabla \boldsymbol{x}'_n|)}$
\State $\boldsymbol{x}_{n+1} \gets \boldsymbol{x}_n - \tau \boldsymbol{A}^T p_{n+1} + \tau divq_{n+1}$
\State $\boldsymbol{x}'_{n+1} \gets \boldsymbol{x}_{n+1} + \theta(\boldsymbol{x}_{n+1} - \boldsymbol{x}_n)$
\State $n \gets n + 1$
\Until $n \geq N$
\State $x_i \gets x'_{n+1}$ \Comment{After $N$ PDHG Iterations}
\EndFor
\State \textbf{return} $x_0$
\end{algorithmic}
\end{algorithm}

\begin{figure*}[!t]
\centering
\includegraphics[width=\textwidth]{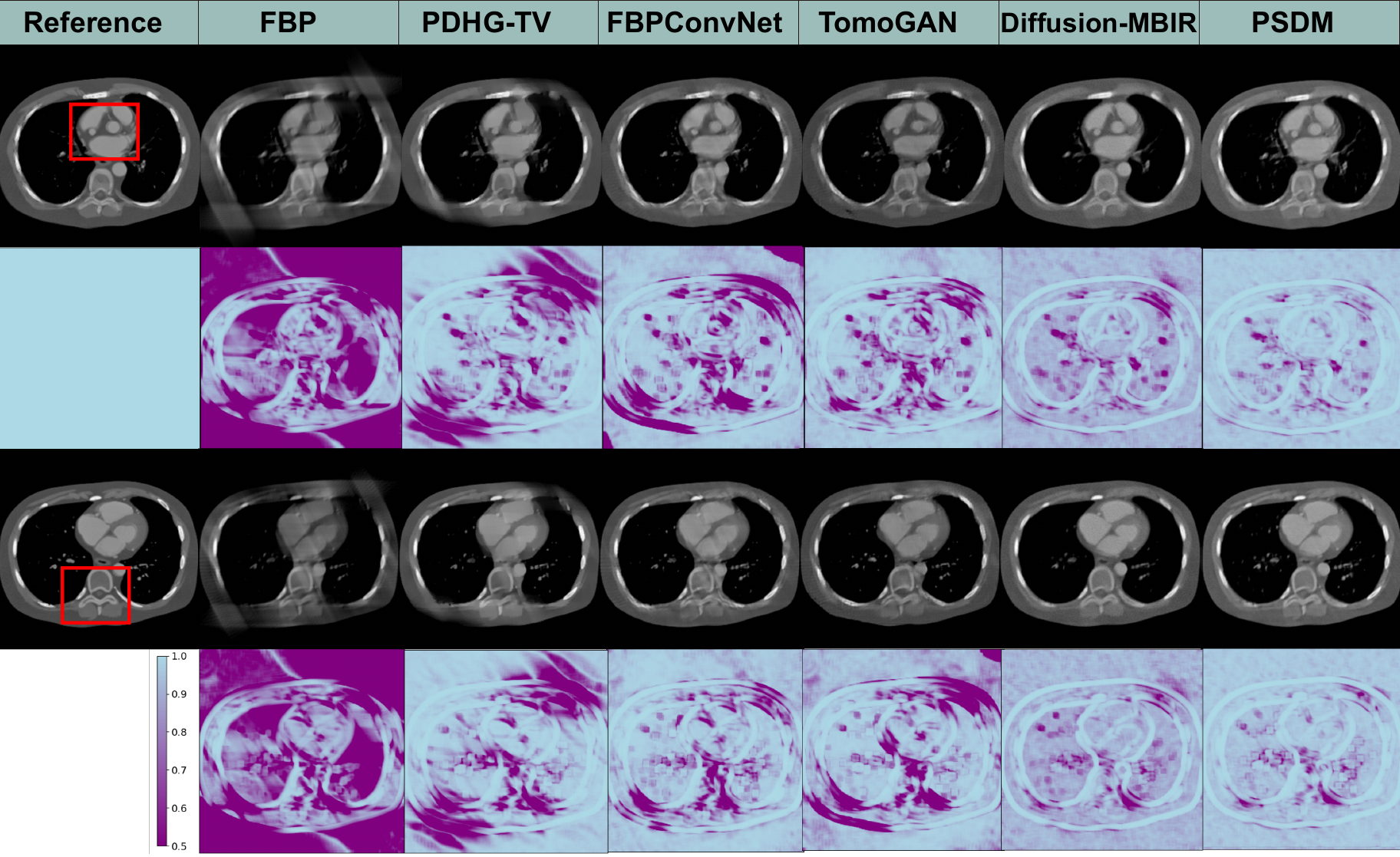}
\caption{Reconstruction results from the simulated dataset for different methods with a scanning angular range of 120° with Case 1 and Case 2. The $2^{nd}$ and $4^{th}$ rows show the SSIM map compared to the ground truths. The display window is [$-540$ $1000$] HU.}
\label{120viewResult1}
\end{figure*}

\section{Experiments and results}
In this section, we first provide an overview of our cardiac CT simulation dataset and outline the essential steps for the training and evaluation of the proposed method. We then compare our results with those obtained from Diffusion-MBIR, another state-of-the-art diffusion model used for CT reconstruction. We also conduct a comparison against several fully-supervised methods. Specifically, we use TomoGAN (a GAN-based generative model) and FBP-ConvNet as our baselines. Finally, we include PDHG-TV, which uses a regularization function, in our study.
\subsection{Numerical simulations}
\subsubsection{Dataset preparation and network training}
We use the 4D Extended Cardiac-Torso (XCAT) phantom version 2 to generate cardiac CT phantom data. XCAT is a multimodal phantom developed at the Duke University \cite{segars20104d,xu2022cardiac}. XCAT can produce a 4D attenuation coefficient phantom to simulate a patient with a beating heart to generate a dynamic cardiac CT scan. We empirically calibrate the tube current-time product (mAs) to ensure that the noise level of the simulated CT images matches that of the real cardiac CT images. In our study, we set mAs per projection to 0.25. And we conceptualize a single cardiac cycle as a duration of one second. Within this framework, we divide each cardiac cycle into 200 discrete phases and sample at equal time intervals to ensure precise temporal resolution. Utilizing these parameters, we generate simulated images for 10 individuals and reconstruct images at phases ranging from 20\% to 80\% with an interval of 12\%. We employ an equiangular cone beam projection geometry, which is \textcolor{red}{based on a model of a GE} CT scanner. The detector array is comprised of 835 units, while the complete scan encompasses 984 views. The field-of-view (FOV) diameter covers an expansive 50 × 50 \textcolor{red}{$\mathrm{cm^2}$}. The distance between the X-ray source and origin \textcolor{red}{is set to 53.852 $\mathrm{cm}$}.  In this study, a cross-validation approach is employed to train our model based on the noise conditional score network plus plus(\textbf{NCSNPP}) \cite{song2020sde}. The database comprises 10 patient datasets, each containing 2D slices of 256 × 256 pixels, \textcolor{red}{representing} a total of 3,714 slices. In each fold of the cross-validation, nine patient datasets (3,362 slices) are used to train the model, while the remaining patient dataset (352 slices) is used for testing. This process is repeated 10 times, with each patient dataset being used exactly once as the testing set. Specifically, in the simulation results section, phase100 in patient 153 is labeled as Case 1, phase112 is labeled as Case 2.

For inference, we adopt the PC sampling strategy with $I$=1,000. In every iteration, data consistency is maintained through the application of the PDHG algorithm, utilizing a sequence of 30 iterations. The experiments are conducted on a high-performance workstation equipped with an Intel i9-9920X CPU operating at 3.50 GHz and an NVIDIA GeForce 2080TI graphics processing unit (GPU). The proposed method is implemented in the PyTorch framework and optimized using the Adam optimizer. The batch size is set as 1, and the network is trained for 200 epochs. It is noteworthy that the Operator Discretization Library (ODL)\cite{adler2017odl} is utilized to implement the PDHG algorithm and the Astra toolbox is adopted to facilitate fast forward and backward projections for image reconstruction in X-ray tomography.

\subsubsection{Simulation Results}
Fig. \ref{120viewResult1} shows the representative reconstruction results for patient 153 across phases 100 and 112, utilizing different methods for 120-degree limited-angle reconstruction. The first and third rows display the reconstruction results, while the second and fourth rows provide the corresponding SSIM map for comparative analysis. For the 120-degree condition, most of the methods achieve satisfactory results, except for FBP and PDHG-TV, which are affected by strong limited-angle artifacts.
\begin{figure}[!t]
  \begin{minipage}{{\columnwidth}}
    \centering
    \includegraphics[width=\linewidth]{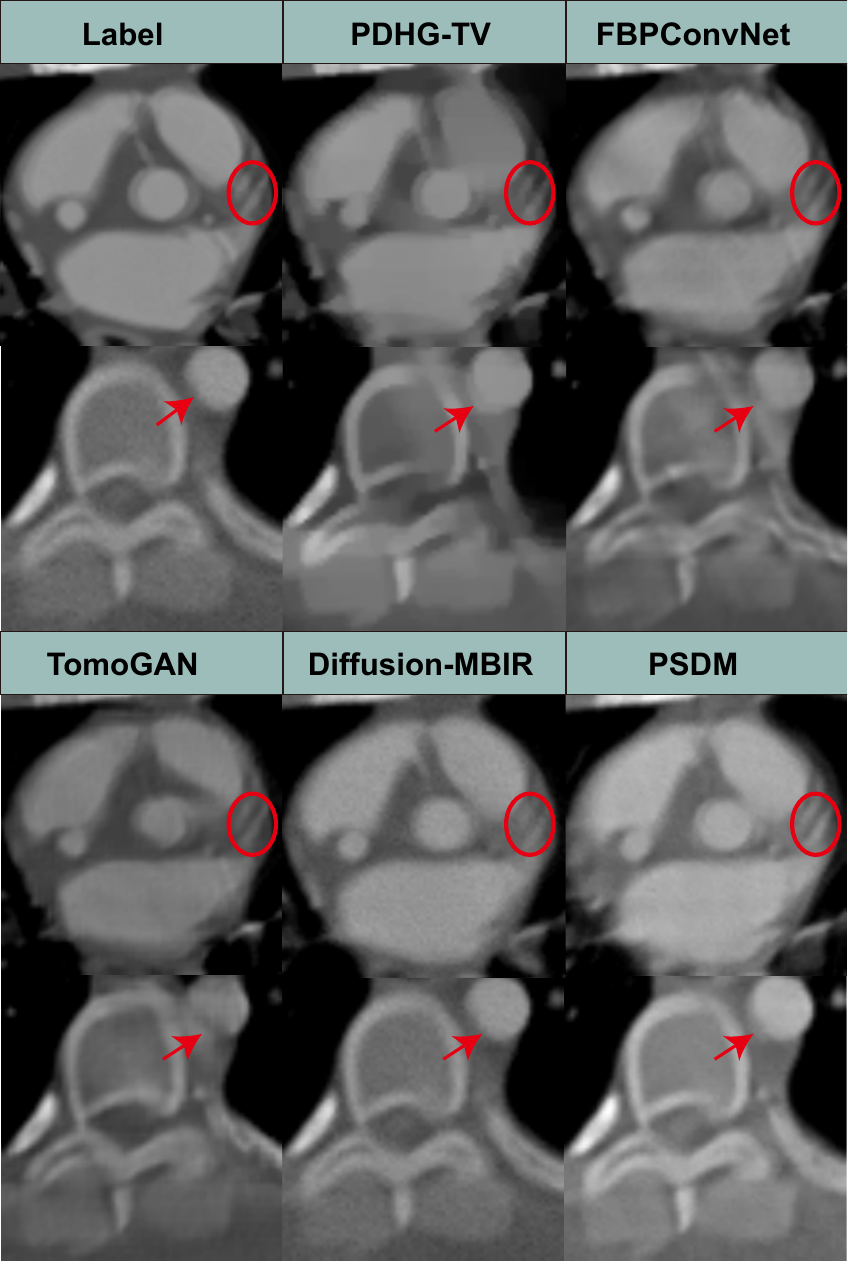}
    \caption{Magnified region of interests for different methods with a scanning angular range of 120° for Case 1 and Case 2. The magnified areas are marked in Fig. \ref{120viewResult1}. The display window is [$-540$ $1000$] HU.}
    \label{120viewResultROI}
  \end{minipage} 
\end{figure}

\begin{figure}[h]
  \begin{minipage}{0.5\textwidth}
    \centering
    \includegraphics[width=\linewidth]{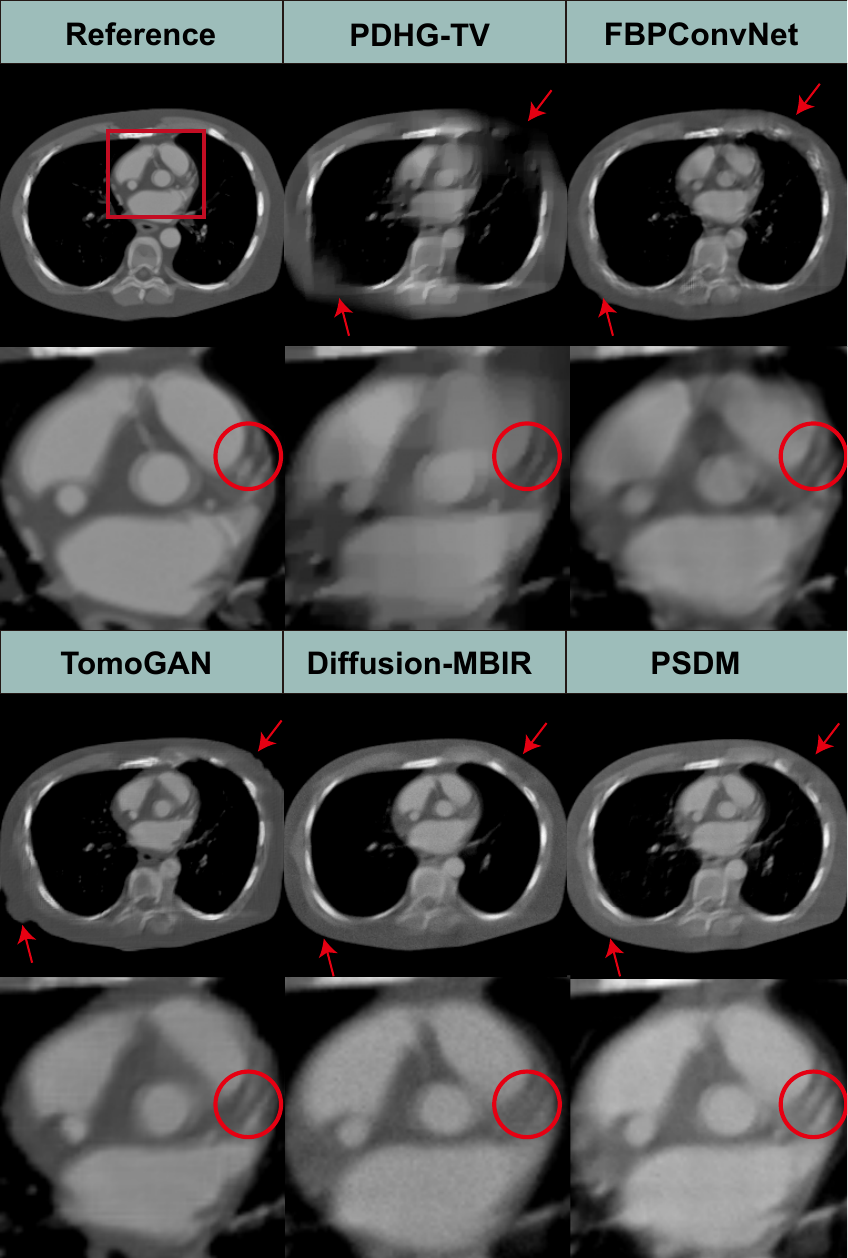}
    \caption{Reconstruction results from simulated projections for different methods with a scanning angular range of 90° for Case 1. The $2^{nd}$ and $4^{th}$ rows show the magnified ROI of the original image. The display window is [$-540$ $1000$] HU.}
    \label{90viewResultROI}
  \end{minipage} 
\end{figure}
FBPConvNet and TomoGAN show compromised edge quality \textcolor{red}{along oblique directions}, while the Diffusion-MBIR and PSDM demonstrate superior edge preservation. However, the Diffusion-MBIR images lack the finer details that are perceivable in PSDM images.

For further comparison, two regions of interest (ROIs) are extracted from Cases 1 and Case 2, and subsequently magnified as in Fig. \ref{120viewResultROI}. Upon close examination of the structures, specifically those indicated by the red arrows, it is evident that the images are subject to blur and distortion when they are processed by PDHG-TV, FBPConvNet, and TomoGAN. Conversely, the Diffusion-MBIR and PSDM yield superior image quality. However, it should be noted that Diffusion-MBIR images are not entirely devoid of imperfections, as they continue to retain some noise and exhibit minor distortions at the locations indicated by the red arrows. Furthermore, when looking closely at the area indicated by the red circles, PSDM provides clearer reconstruction of the LAD branches compared to Diffusion-MBIR.

\begin{table*}[!t] \tiny
\caption{Quantitative evaluation of XCAT testing data. \textbf{Bold}: Best, \underline{under}: second best.}
\centering
\setlength{\tabcolsep}{0.2em}
\resizebox{0.7\textwidth}{!}{%
\begin{tabular}{lcccccccc}
\toprule
{} & \multicolumn{4}{c}{\textbf{120-view}}  & \multicolumn{4}{c}{\textbf{90-view}}\\
\cmidrule(lr){2-5}
\cmidrule(lr){6-9}
\textbf{Method} & PSNR $\uparrow$ & SSIM $\uparrow$ & HC $\uparrow$ & LBP-TS $\downarrow$ & PSNR $\uparrow$ & SSIM $\uparrow$ & HC $\uparrow$ & LBP-TS $\downarrow$\\
\midrule
PDHG-TV\cite{sidky2012convex} & 27.856 & 0.860 & 0.911 & 0.191 & 24.067 & 0.765 & 0.498 & 0.201\\
FBPConvNet\cite{jin2017deep} & 30.552 & \underline{0.912} & \underline{0.927} & 0.023 & \underline{27.82} & \underline{0.870} & \underline{0.821} & 0.008 \\
TomoGAN\cite{liu2020tomogan} & 24.698 & 0.854 & 0.781 & \underline{0.009} & 22.861 & 0.818 & 0.813 & \underline{0.007}\\
Diffusion-MBIR\cite{chung2023solving} & \underline{32.991} & 0.902 & 0.797 & 0.011 & 27.01 & 0.827 & 0.743 & 0.01\\
PSDM (ours) & \textbf{35.56} & \textbf{0.938} & \textbf{0.942} & \textbf{0.0008} & \textbf{30.60} & \textbf{0.901} & \textbf{0.912} & \textbf{0.0015}\\
\bottomrule
\end{tabular}
}
\label{quantitative}
\end{table*}

We conduct further tests using a 90-degree limited-angle dataset. The results, along with the magnified ROI images, are presented in Fig. \ref{90viewResultROI}. We observe that the PDHG-TV method exhibits noticeable limited-angle artifacts. Both  FBPConvNet and TomoGAN demonstrate varying degrees of deformation at the diagonal edges. The diffusion-based methods appear to be more stable, without significant evident structural deformation. \textcolor{red}{Similar to the 120 degree limited-angle reconstruction results, our PSDM method outperforms the Diffusion-MBIR in reconstructing finer structural details in LAD area, as highlighted by the red circles.}

\subsubsection{Quantitative Evaluation}
A quantitative assessment is performed for different competing reconstruction techniques. In addition to the commonly used SSIM and PSNR, we also employ histogram correlation (HC) and local-binary-pattern-based texture similarity (LBP-TS) as evaluation metrics. HC measures the similarity between the histograms of reconstruction results and labels. A high correlation indicates that the images have similar intensity distributions, which is crucial for tasks like CT reconstruction where maintaining intensity distribution is desirable. LBP-TS evaluates the texture similarity between two images by extracting local binary patterns from them and comparing the patterns\cite{ojala2002LBP}; smaller values indicate higher textural image similarity.  

The images that are reconstructed in two different scenarios (120\textdegree  and 90\textdegree) are analyzed, and the quantitative metrics are presented in Table \ref{quantitative}. Consistent with visual impressions, PSDM yields more favorable quantitative outcomes compared to the PDHG-TV, FBPConvNet, TomoGAN, and Diffusion-MBIR methods. In particular, the PSDM obtains the highest value of PSNR, SSIM, and HC in all cases. In terms of LBP-TS, the PSDM consistently achieves the smallest (most desirable) values. The empirical findings from this quantitative analysis provide a substantive validation of the merits of the PSDM method, underline its superior performance for limited-angle image reconstruction under the current test conditions.

\subsection{Experiments on clinical data}
For cardiac CT imaging, it is crucial to freeze the beating heart \textcolor{red}{to avoid} motion artifacts. However, the conventional ECG-gating does not work well for patients with tachycardia and arrhythmias, and radiation exposure is relatively high due to the overlapped continuous scanning needed for retrospective gating. By using limited-angle reconstruction techniques, temporal resolution can be significantly improved, allowing for clearer visualization of fast-moving structures such as heart valves and myocardial walls. 
\subsubsection{Stanford AIMI COCA results}
\begin{figure}[h]
  \begin{minipage}{0.5\textwidth}
    \centering
    \includegraphics[width=\linewidth]{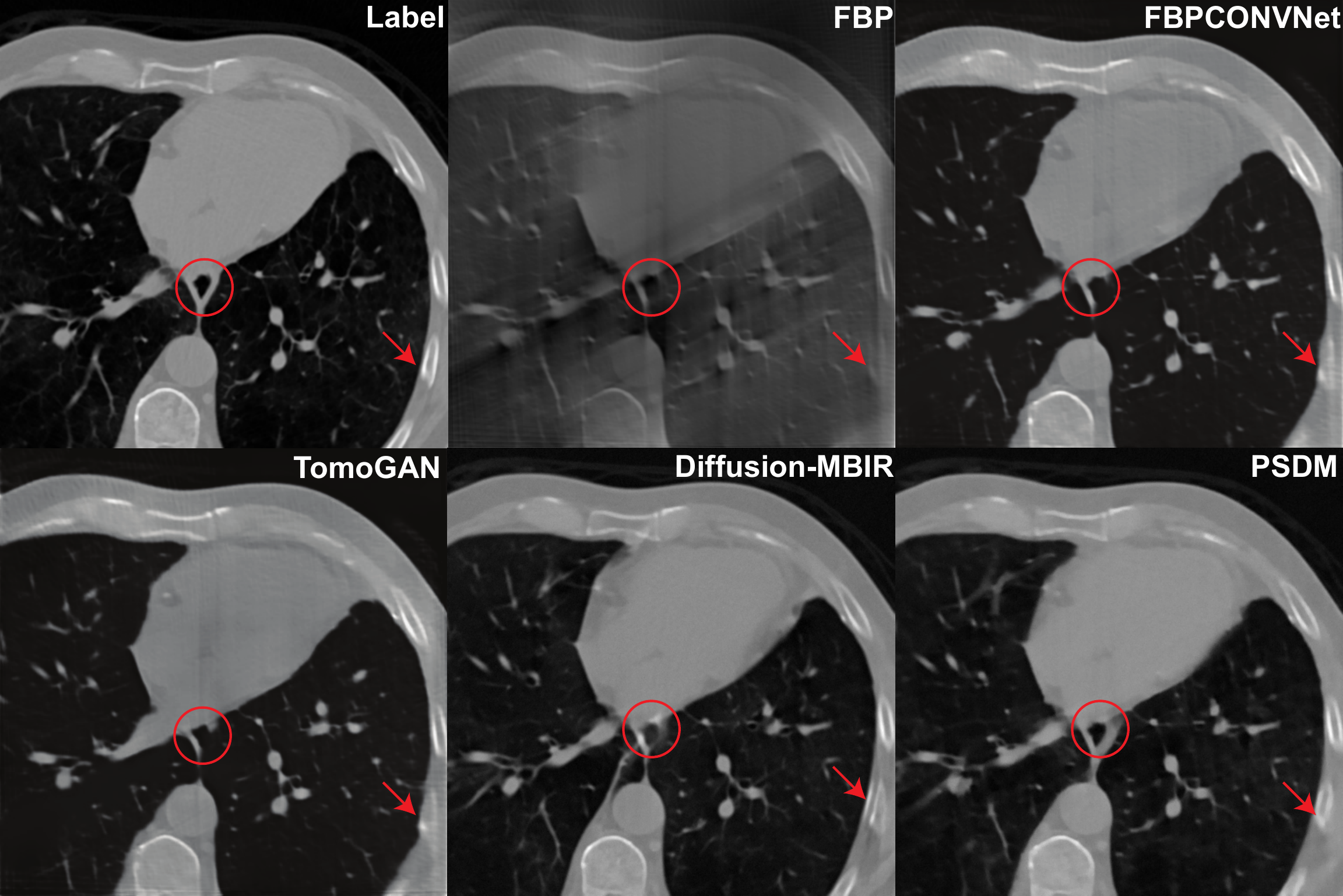}
    \caption{Reconstruction results from the AIMI COCA dataset with 120 views using different methods. The display window is [$-900$ $1000$] HU.}
    \label{Stanford}

  \end{minipage} 
\end{figure}

To advance our clinical dataset evaluation, the Stanford AIMI COCA dataset\cite{StanfordAIMI2021} is employed. The COCA dataset \textcolor{red}{is a compilation} of 789 gated CT scans, with each scan corresponding to an individual patient. On average, every patient contributes roughly 50 image slices, and each image slice possesses a resolution of 512 x 512 pixels. The NCSNPP model is trained with data from 60 patients, totaling approximately 3,400 image slices. Apart from the image size, all other parameters remain consistent with the simulation training. The reconstruction results from 120 views \textcolor{red}{appear} in Fig. \ref{Stanford}. Significantly, our proposed method produces images of enhanced quality characterized by prominent features and sharp image boundaries. This is especially discernible in the areas and structures underscored by red circles and arrows. In contrast, other algorithms tend to obscure certain details and introduce blurring.

\subsubsection{Clinical cardiac results}
\begin{figure}[!t]
  \begin{minipage}{0.5\textwidth}
    \centering
    \includegraphics[width=\linewidth]{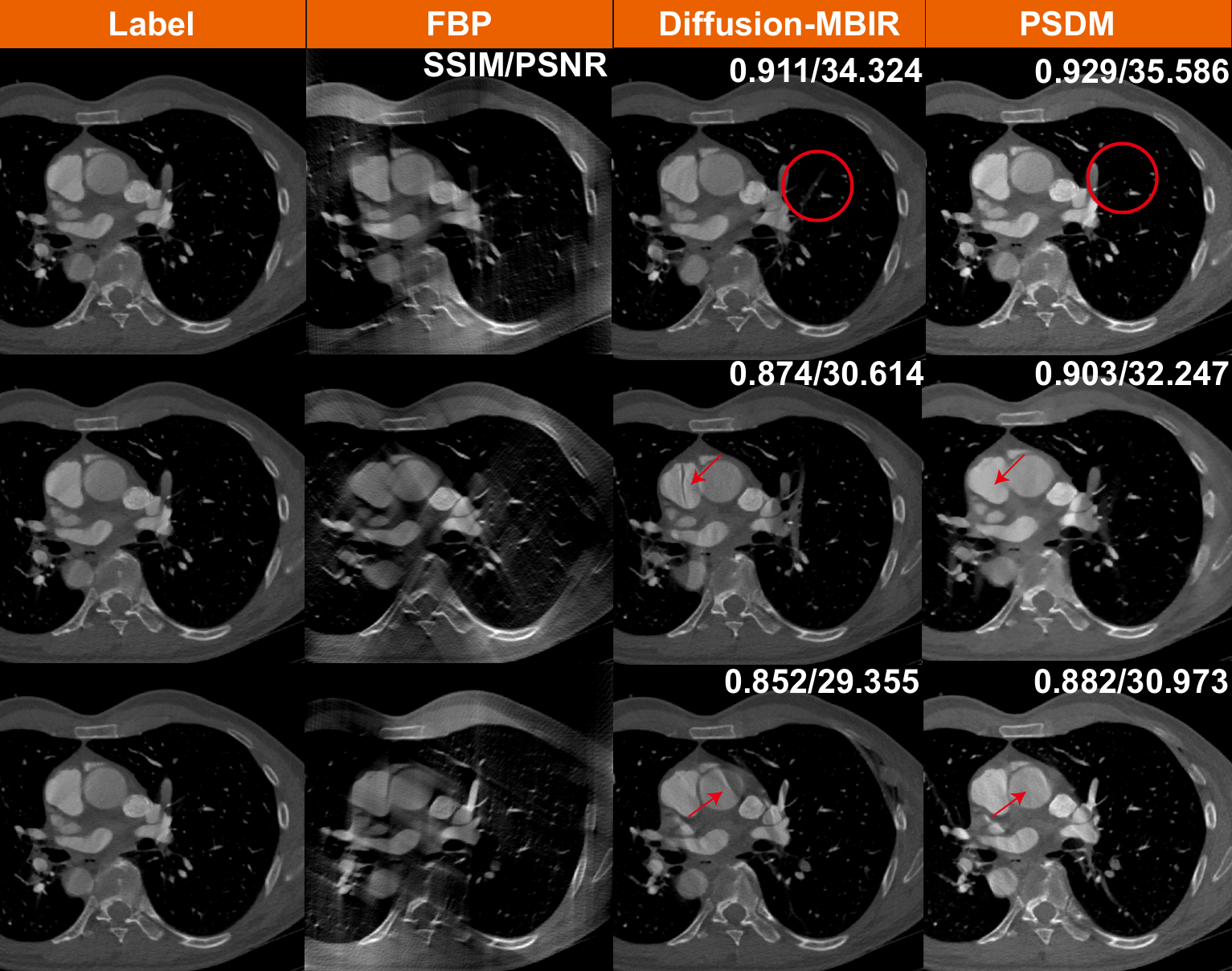}
    \caption{Reconstruction results from a clinical cardiac CT scan, using different methods, each corresponding to different scanning angular ranges of 120°. The first row are reconstructions from spanning angles 0° to 120°, the second row are for 30° to 150°, and the third row are for 60° to 180°. The display window is [$-400$ $1500$] HU.}
    \label{real}

  \end{minipage} 
\end{figure}

Generalization is a major issue for DLR-based methods, especially for models requiring supervised learning. If a model is trained strictly on data with a specific angle range (\emph{e.g.}, 0--120\textdegree), it will not generalize well to other angular ranges. However, different clinical situations might require different angular ranges. For instance, certain anatomical regions or pathologies are better visualized with a specific angle range.  To further demonstrate the advantages of PSDM, the algorithm is applied to a real clinical cardiac dataset. With the approval of the Institutional Review Board of the University of Massachusetts, Lowell, a deidentified clinical cardiac CT dataset is obtained from a GE HD 750 scanner. \textcolor{red}{The patient was scanned using axial mode.} 1,520 projections are acquired over an angular range of 556\textdegree \ ($\approx$ 1.54 rotations). Each projection row has 835 elements at 1.095 mm pitch. The source to the rotation center distance is 538.5 mm, and the source to the detector distance is 946.7 mm. Reconstruction is performed by FBP using equi-angular geometry, where the image size is set to $512\times512$ to accelerate the sampling process. Since the supervised learning-trained models (FBPConvNet and TomoGAN) usually can only process specific angle range,  we compare only the two unsupervised learning methods: Diffusion-MBIR and PSDM. During the sampling stage, these two methods directly utilize the checkpoints trained on the aforementioned Stanford AIMI COCA datasets. Fig. \ref{real} shows the 120-view limited-angle reconstruction images from different angular ranges. We find that both Diffusion-MBIR and PSDM produce consistent and apparently acceptable results. However, compared with the Diffusion-MBIR, the results produced by PSDM exhibit fewer limited-angle artifacts and appear to provide a more realistic representation. These differences can be seen in the areas and structures underscored by red circles and arrows. Specifically, in the atrial region denoted by the red arrow, PSDM produces more stable results, while Diffusion-MBIR still exhibits limited-angle artifacts in the atria.

\subsection{Ablation study}
In the ablation study section, we explore the impact of various components and parameters in our proposed method, mainly focuses on three aspects: the number of reverse diffusion steps, the effect of the TV term, and how Fourier fusion module affect the denoising process. This examination is crucial to understand the contribution of each element to the overall performance of the system.
\begin{figure}[!t]
  \begin{minipage}{0.5\textwidth}
    \centering
    \includegraphics[width=\linewidth]{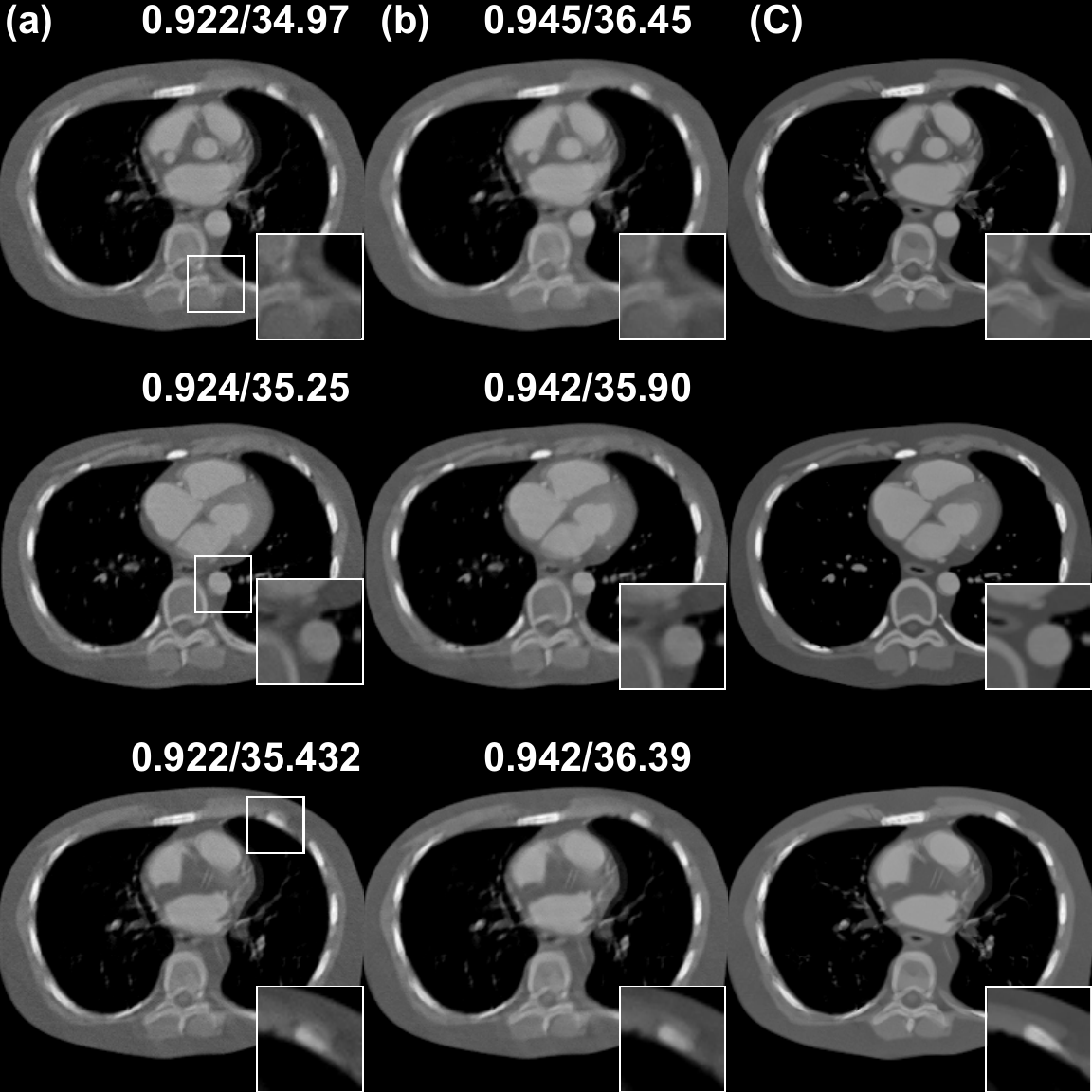}
    \caption{Ablation study for the effect of TV term with 120 views using PSDM. (a) w/o TV prior, (b) w/ TV prior, (c) Ground truth. The display window is [$-540$ $1000$] HU. }
    \label{ab2}
  \end{minipage} 
\end{figure}

\begin{figure}[!t]
    \centering
    \includegraphics[width=\linewidth]{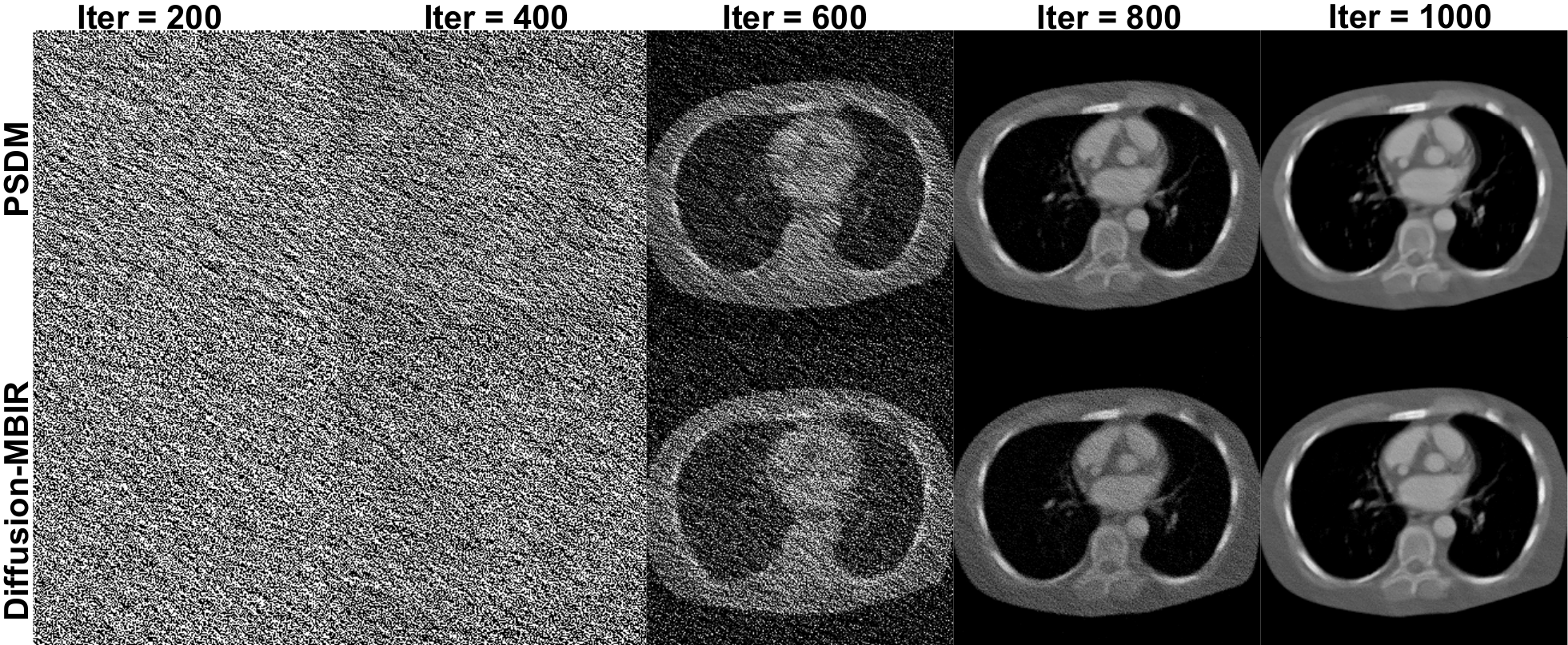}
    \caption{Reconstructed images of the PSDM and Diffusion-MBIR with respect to different reverse diffusion steps on 120 views. The display window is [$-540$ $1000$] HU.}
    \label{ab1}
\end{figure}
Figure. \ref{ab2} presents a comparative analysis of the reconstruction results with and without the TV term. These quantitative metrics (SSIM and the PSNR) indicate that the TV term can improve the perceptual quality and the fidelity of the reconstructed images. Visually, results with the TV term exhibit less noise and more consistent structural integrity compared to those without the TV prior. Figure. \ref{ab1} illustrates the performance comparison between PSDM and Diffusion-MBIR with respect to numbers of reverse diffusion steps. It is observed that PSDM demonstrates faster convergence relative to Diffusion-MBIR. Specifically, in the iterations ranging from 600 to 800, Diffusion-MBIR exhibits higher noise levels compared to PSDM. Figure \ref{FF} provides a quantitative comparison of the performance with and without the Fourier fusion module. It is observed that between 400-800 reverse steps, the PSDM (presumably a metric being compared) with the Fourier fusion module outperforms the one without it. Ultimately, both the PSNR and SSIM metrics show a slight improvement when the Fourier fusion module is utilized, highlighting its benefits.

\begin{figure}[!t]
  \begin{minipage}{0.5\textwidth}
    \centering
    \includegraphics[width=\linewidth]{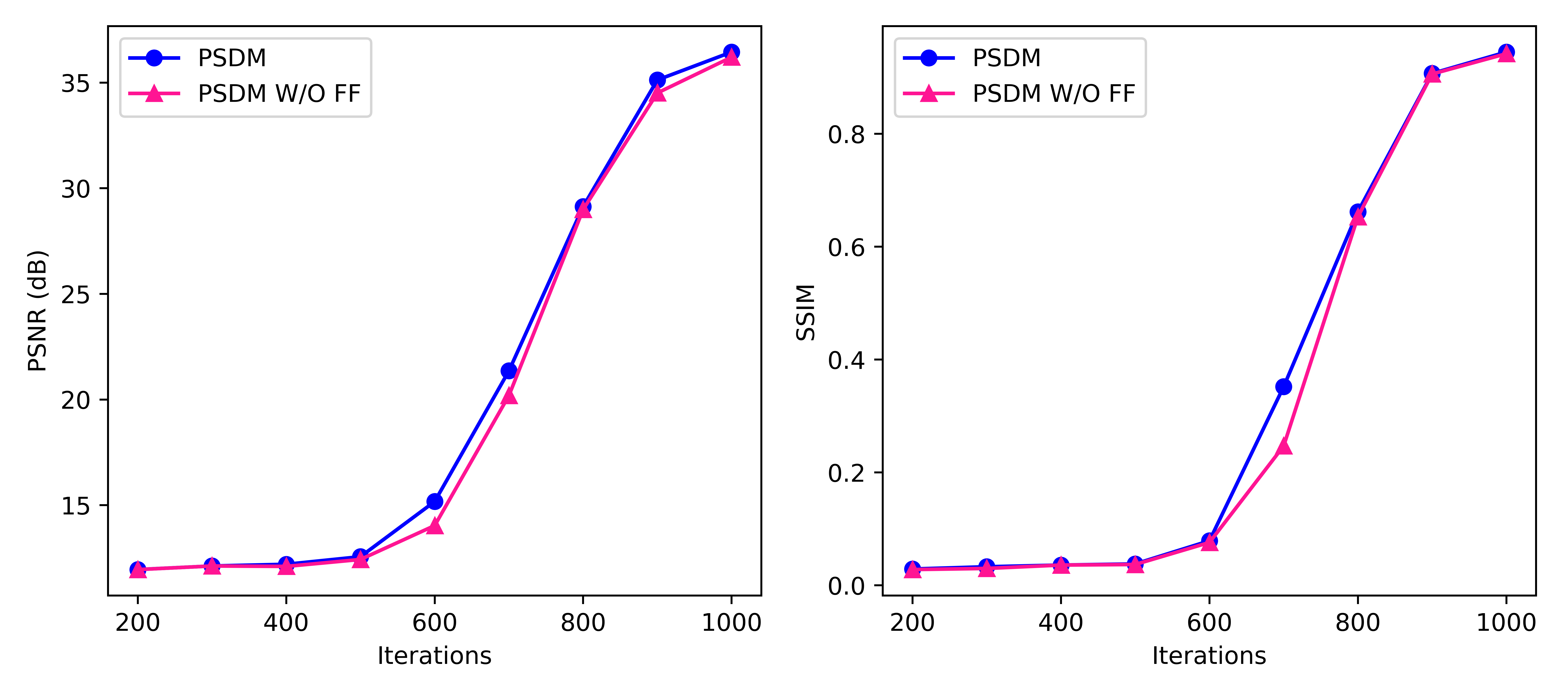}
    \caption{Ablation study for the effect of FF module with 120 views using PSDM.}
    \label{FF}
  \end{minipage} 
\end{figure}

\section{Discussion \& Conclusion}
In the realm of language models, we have witnessed the rise of models such as ChatGPT, which have achieved impressive generalizability. Similarly, the next leap in imaging might be a generalized image model, versatile across tasks and less reliant on specific datasets. While current computational imaging methods merge traditional and modern data-driven techniques, they often falter in unfamiliar conditions due to dataset dependencies. This highlights the need for diverse datasets that are hard to procure in areas \textcolor{red}{such as} cardiac CT. The score-based diffusion model offers a solution. It can generate new samples from a learned data distribution and without requiring a network architecture specific to the problem at hand. Usually a standard architecture such as U-Net can be employed as denoiser, and adversarial learning is not required. Thus, diffusion models fill an essential gap in the landscape of generative models, offering an efficient and uncomplicated approach to model complex data distributions. Such streamlined design, combined with a straightforward training process, offers a powerful alternative to models that require either intricate design, or pose complex adversarial training challenges. As an unsupervised deep learning model, it is adaptable and less dataset-dependent. By incorporating physics-based methods, it emphasizes understanding physical processes rather than data reliance, broadening its conditions range and enhancing generalizability.

The combination of PDHG and SDE for CT reconstruction in this area suggests that PDHG could potentially be replaced with another model-based reconstruction algorithm. However, PDHG known for its effectiveness in solving non-smooth optimization problems, particularly in image processing, might have a specific synergy with the SDE-based approach. For instance, the primal-dual nature of PDHG could be especially 
adept at handling the noise model or the data fidelity term as formulated in the SDE approach. Secondly, PDHG is known for its strong convergence properties, especially in the context of convex optimization problems with non-smooth terms. If SDE-based denoising leads to an optimization landscape that benefits from these convergence properties, this would be a significant justification for using PDHG over other algorithms.

\begin{figure}[!t]
  \begin{minipage}{0.5\textwidth}
    \centering
    \includegraphics[width=\linewidth]{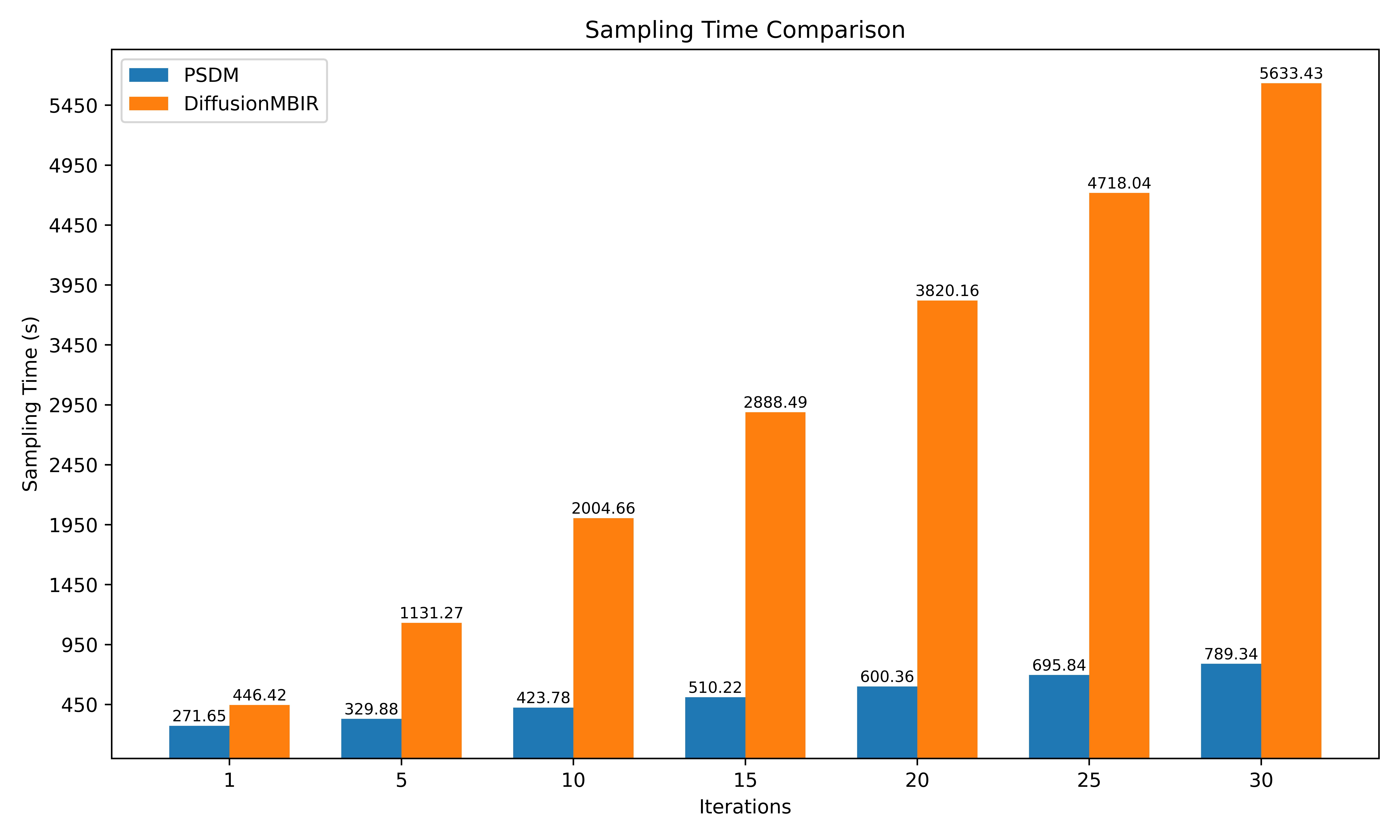}
    \caption{Comparison of sampling time between the PSDM and Diffusion-MBIR with respect to different iteration numbers (1--30) on XCAT dataset.}
    \label{sample time}
  \end{minipage} 
\end{figure}

\begin{figure}[!t]
  \begin{minipage}{0.5\textwidth}
    \centering
    \includegraphics[width=\linewidth]{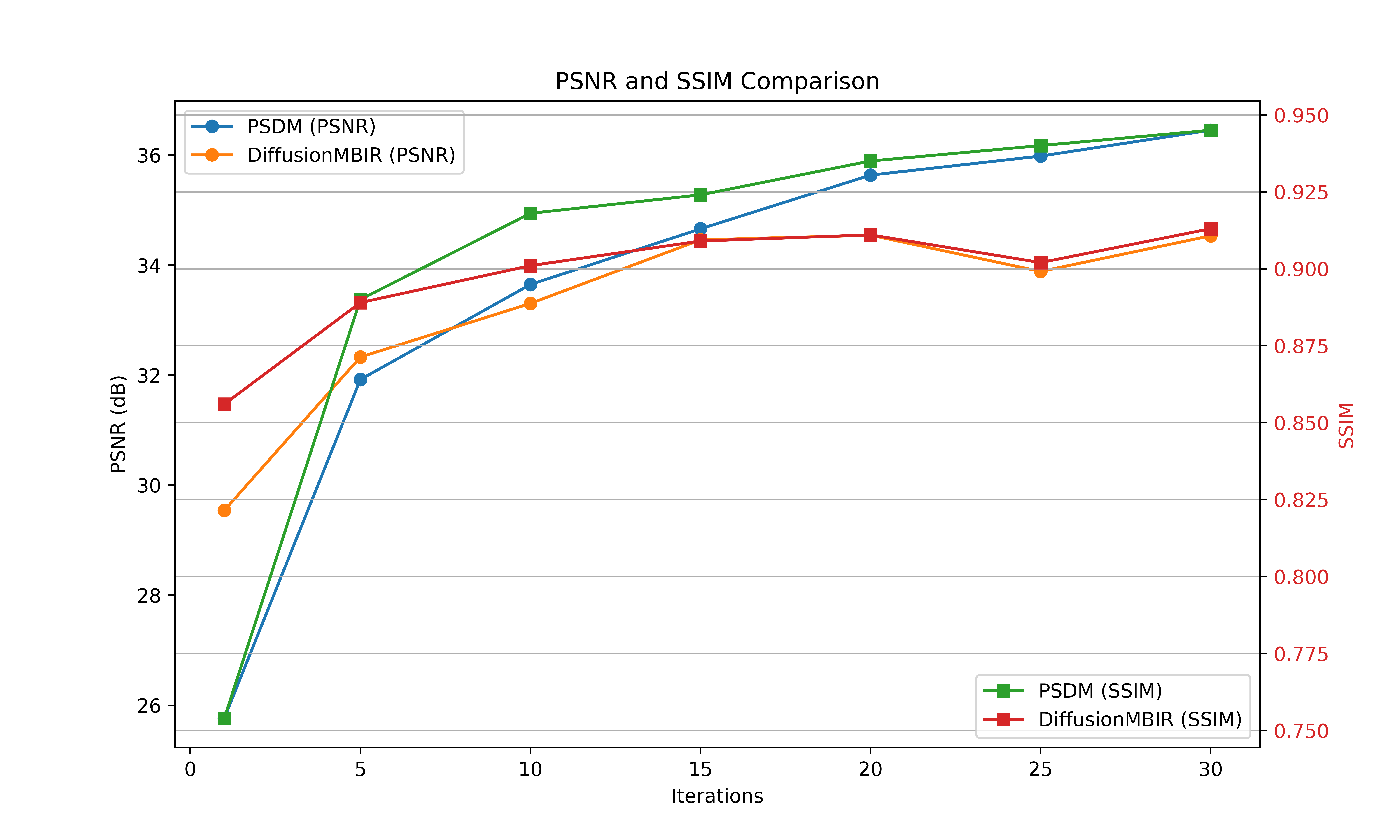}
    \caption{Performance analysis of the PSDM and Diffusion-MBIR in terms of PSNR and SSIM values with respect to the iteration numbers(1--30) on patient 153 with 120 views.}
    \label{psnr}
  \end{minipage} 
\end{figure}

To prove the advantage of PDHG, a set of experiments are conducted to evaluate the importance of the PDHG in PSDM. The major difference between PSDM and Diffusion-MBIR is the data consistency component. For the PSDM, we employ the PDHG algorithm, which has demonstrated superior effectiveness and robustness compared to the ADMM algorithm used in Diffusion-MBIR. It is well known that the number of iterations is an important parameter in optimization algorithms. Typically, increasing the number of iterations improves the final results, but this comes at the cost of a significant rise in computational time. Fig. \ref{sample time} presents a comparison between PSDM and Diffusion-MBIR as a function of the number of iterations, sampling one image. It is evident that as the iteration number increases, the computational cost of Diffusion-MBIR increases dramatically. In contrast, the sampling time for the proposed PSDM remains relatively stable and acceptable. Furthermore, insights from Fig. \ref{psnr} reveal that, initially, the Diffusion-MBIR achieves superior PSNR and SSIM values compared to the PSDM. However, as the iteration number grows, PSDM witnesses a significant improvement in both PSNR and SSIM. On the other hand, the Diffusion-MBIR appears to plateau, suggesting a performance ceiling.

However, while PSDM \textcolor{red}{exhibits} a significantly reduced sampling time compared to the Diffusion-MBIR method, it falls short in applications requiring real-time deartifacting. This is a crucial aspect that needs to be addressed, because accelerating the sampling procedure could greatly enhance the model's applicability in time-sensitive tasks, such as in a clinical setting where quick decision-making is essential. \textcolor{red}{While our proposed approach is empirically driven, we continue to investigate the supporting theory. Although} PSDM shows good performance in handling a variety of datasets, its adaptability has limits \textcolor{red}{and there is room to further improve image quality.} This indicates a need for further refinement in its ability to generalize across more diverse data types. Theoretically speaking, it is not ideal to apply projection steps to the noisy variable in both Diffusion-MBIR and PSDM because the noisy variable does not belong to the subspace of the system matrix A due to the added noise. In the future, we plan to adopt a more accurate probabilistic formula such as DPS \cite{chung2022posterior}. DPS estimates mean of $x_0$ from $x_t$ and apply the data fidelity to the estimated $x_0$. However, it introduces a Jacobian term, \textcolor{red}{and a hard consistency method is introduced in some papers to avoid this}\cite{song2023hard}. We will consider incorporating this approach in the future. 

In conclusion, while data-driven methods have already revolutionized the field of imaging and computer vision, the limitations of these methods in terms of generalizability and data dependency cannot be overlooked. The physics-informed score-based diffusion model, by synergizing data- and model-based methods, offers a promising direction, emphasizing the importance of understanding the underlying physics. This approach not only mitigates the challenges posed by data dependency but also broadens the horizons for tackling diverse conditions in cardiac CT reconstruction and without a significant increase in computational cost.

\bibliography{references}
\bibliographystyle{IEEEtran}

\end{document}